\documentclass[a4paper,11pt]{article}

\usepackage[dvips]{graphicx}
\usepackage{epsfig}
\usepackage{pstricks}
\usepackage{amsmath,amssymb,amsfonts}
\usepackage{color}
\usepackage{xspace}
\usepackage{a4wide,epic}
\usepackage{lipsum}
\usepackage{mathtools}
\usepackage{cuted}
\usepackage{multicol}
\usepackage[normalem]{ulem}

\newcommand{\tsz}{\tilde{\sigma}_z}
\newcommand{\sz}{\sigma_z}
\newcommand{\ba}{\mathbf{a}}

\newcommand{\q}[1]{|#1\rangle}
\newcommand{\qd}[1]{\langle #1|}

\begin{document}

\title{Quantum adiabatic elimination at arbitrary order for
photon number measurement\footnote{This work has been supported by ANR grant HAMROQS.}}

\author{Alain Sarlette, Pierre Rouchon, Antoine Essig, Quentin Ficheux and Benjamin Huard
\footnote{A.S.~is with QUANTIC lab, INRIA Paris and ELIS department, UGent. P.R.~is with Centre Automatique et  Syst\`{e}mes, Mines-ParisTech. A.E. and B.H. are with the Physics Laboratory at ENS Lyon. Q.F. is with the Joint Quantum Institute, University of Maryland.}}

\maketitle

\begin{abstract}
Adiabatic elimination is a perturbative model reduction technique based on timescale separation and often used to simplify the description of composite quantum systems. We here analyze a quantum experiment where the perturbative expansion can be carried out to arbitrary order, such that: (i) we can formulate in the end an exact reduced model in quantum form; (ii) as the series provides accuracy for ever larger parameter values, we can discard any condition on the timescale separation, thereby analyzing the intermediate regime where the actual experiment is performing best; (iii) we can clarify the role of some gauge degrees of freedom in this model reduction technique.
\end{abstract}


\section{Introduction}

Model reduction is an ubiquitous way to make system analysis and design more tractable. This is especially relevant in quantum systems, where the dimension of the full state is the product of the dimensions of its components' states (\cite{NielsenChuang}). Adiabatic elimination approaches model reduction via timescale separation: the transients associated to fast degrees of freedom are discarded in order to write a lower-dimensional model describing the slow degrees of freedom. In principle, as confirmed by center manifold theory (\cite{Fenichel79}), there exist exact invariant subspaces for the slow dynamics, but in general they are hard to compute. Perturbative expansions as a function of the timescale separation can provide approximate slow-variable models at various orders, with reasonable computational effort (\cite{carr-book}). In quantum physics, the lowest-order approximation is routinely used to summarize the effects on a target system, of standard couplings to fast surrounding components -- in physicists' words, induced dephasing and dissipation/broadening of resonance peaks as a perturbation, or reservoir engineering power when the induced dissipation/stabilization is the very goal (\cite{HarocheBook}). As quantum control is reaching for more precision, higher accuracy approximations become of interest.

During the recent years, the technique of center manifolds and perturbative expansion have been systematically adapted to composite quantum systems (\cite{azouitQST2017,bouten-silberfarb-08,zanardi2016dissipative}). General formulas have been obtained for the first and second order approximation. Besides the actual computations, there are two main points. The first one is that the quantum master equation is linear, so in principle invariant subspaces should exist and be analytic in the expansion parameter. The second point however, is that quantum systems must have a particular structure: they describe the completely positive evolution of a positive semi-definite state matrix of trace one (\cite{HarocheBook}). This structure is the basis to predict any probabilistic output from the system, so the reduced model can be physically interpreted as such only if it takes this particular structure too. It has been shown how this is obtained systematically in first and second order approximation (\cite{azouitQST2017}), but not at higher order yet.

In this paper, we treat a case where adiabatic elimination can be carried out to arbitrary order with low computational effort, and we prove how the completely positive structure can be preserved all the way. The system under study comes from an actual experimental setup for measuring the photon number in a harmonic oscillator (\cite{ExpPaper2019}).

In this experiment, carried out at ENS Lyon, the photon number in a harmonic oscillator (HO -- playing the role of the slow subsystem) is measured continuously and without destroying the photons, thanks to the effect they have on a qubit coupled to the HO. By measuring the light reflected from the qubit (the fast subsystem), information is obtained indirectly and continuously about the HO photon number. This information output implies that in the complementary basis, information is perturbed, at the same rate (\cite{Devoret}). Thus, by studying the slow dynamics governing the perturbation of the HO state, we can deduce how strongly photon number is measured in the setup, and optimize parameters. Non-destructive measurements, harmonic oscillators and qubits are most typical building blocks for quantum hardware (\cite{HarocheBook}), so providing new tools for this case should be practically relevant. Our ability to carry out adiabatic elimination at all orders allows to work in the usually less tractable regime where all timescales are of similar order -- in fact, as we explain below, this is the optimal one.

On the mathematical side, the developments are facilitated by a natural decoupling between degrees of freedom, such that we can identify a structure that carries through at all orders of adiabatic elimination. The results also carry more general lessons about the conjecture that quantum-structure-preserving adiabatic elimination should be feasible at all orders in general quantum systems. First, the conjecture does hold on this example. Second, it is highlighted explicitly how a gauge degree in the slow system parameterization plays a role in obtaining a completely positive model. This may clarify directions and reasonable objectives for addressing more general models.

The paper is organized as follows. Section \ref{sec:model} gives the system model and recalls the adiabatic elimination approach. Section \ref{sec:result} contains our main result, with an analysis of preserving complete positivity in Section \ref{ssec:positivity}. In Section \ref{sec:conseq} we illustrate a few concrete implications that can be computed from the reduced model, and we compare our predictions to experimental results.


\section{Description of the setting}\label{sec:model}

We briefly describe the photon number measurement experiment at ENS Lyon, explaining how it decouples into two-qubit systems which we then study in the next section. The reader is referred to \cite{ExpPaper2019} for more details on the experiment, and to e.g.~\cite{NielsenChuang} for more basic background on quantum systems. More background about our center manifold approach to quantum adiabatic elimination can be found in \cite{azouitQST2017}. We just recall that a quantum state $\rho$ is a positive semi-definite Hermitian operator of trace $1$.


\subsection{The photon number measurement experiment}

The idealized experiment (i.e.~discarding spurious couplings to the environment) can be described in the following mathematical terms. A harmonic oscillator mode (photon annihilation operator $\mathbf{a}$) is coupled dispersively to a possibly detuned and driven qubit (Pauli operators $\sigma_x,\sigma_y,\sigma_z$, spontaneous emission operator $\sigma_-$). 

Owing to this coupling, the reflection of light at a given frequency encodes information about the presence of a given photon number in the HO. Although the full experiment probes the reflection at many frequencies hence many photon numbers simultaneously by multiplexing, for simplicity we will here assume that the drive contains a single frequency. In the interaction frame and after averaging out the fast counter-rotating terms relative to the drive frequency\footnote{Absolute frequencies (GHz for circuit quantum electrodynamics) are much higher than the frequency differences (MHz range)}, the Hamiltonian reads:
$$H = \frac{-\chi}{2} \mathbf{a}^\dagger \mathbf{a} \otimes \sigma_z + \frac{\Delta}{2} \sigma_z + \frac{\Omega}{2} \sigma_x \; .$$
Here $\otimes$ denotes the tensor product between different quantum systems, we drop the tensor with identity $I$, e.g.~$\sigma_z$ means $I \otimes \sigma_z$; $\chi$ denotes the coupling strength ($2\pi 4.9$ MHz in experiment), $\Delta$ is the drive detuning from the qubit frequency when the HO is empty, and $\Omega$ is the drive amplitude (free parameters of the input signal). The qubit state is monitored by continuously detecting its fluorescence field, corresponding to the complex amplitude associated to a measurement channel operator $L = \sqrt{\kappa} \sigma_-$ (detection rate $\kappa = 20$ MHz). 
With the usual superoperator notation $\mathcal{D}_L(\rho) = L \rho L^\dagger - \tfrac{L^\dagger L \rho + \rho L^\dagger L}{2}$ and $\mathcal{M}_L(\rho) = L\rho + \rho L^\dagger - \text{tr}(L\rho + \rho L^\dagger)\rho$, the stochastic master equation for the full system state $\rho$ reads:\vspace{-3mm}
\begin{eqnarray*}
d\rho_t &=& -i[H,\rho_t]\, dt + \kappa\mathcal{D}_{\sigma_-}(\rho_t)\, dt \\
&& + \sqrt{\tfrac{\eta \kappa}{2}} \mathcal{M}_{\sigma_-}(\rho_t) \, dW^{(1)}_t + \sqrt{\tfrac{\eta \kappa}{2}} \mathcal{M}_{i\sigma_-}(\rho_t) \, dW^{(2)}_t
\\
dY^{(1)}_t &=& \sqrt{\eta\kappa/2} \,\text{tr}(\sigma_-\rho_t + \rho_t \sigma_-^\dagger) \, dt + dW^{(1)}_t\\
dY^{(2)}_t &=& \sqrt{\eta\kappa/2} \,\text{tr}(i\sigma_-\rho_t - \rho_t i\sigma_-^\dagger) \, dt + dW^{(2)}_t \; .
\end{eqnarray*}
Here $\text{tr}$ denotes the trace and $\eta \in [0,1]$ is the measurement efficiency. The Wiener processes $dW^{(1)}_t$, $dW^{(2)}_t$ capture the probabilistic character of quantum measurements. The two quadratures of the output signal $(dY_t^{(1)},dY_t^{(2)})$ monitor the qubit state while it is influenced by the harmonic oscillator (HO), therefore providing indirect information about the HO state. As the coupling is through the photon number operator $\mathbf{a}^\dagger\mathbf{a}$, one expects to obtain information about the photon number.

Physicists understand two extreme regimes quite well.
\newline $\bullet$ When the detection rate $\kappa$ (non-unitary effect) is small, a major effect of the coupling is obtained when $\Delta-n\chi=0$ for some integer $n$: the corresponding photon number level $n$ is at resonance while the others are not, and at low drive power $\Omega$ only the resonant level will allow qubit excitations via $\Omega \sigma_x$, and thus spontaneous emissions. In other words, the output amplitude will tell if we are on level $n$ or not. From a mathematical viewpoint, being resonant / off-resonant boils down to an \emph{averaging} approximation or rotating wave approximation (RWA), with well-defined resonances for $\kappa / \chi \ll 1$.
However, when $\kappa$ is small, the system is only able to weakly leak information to us. 
\newline $\bullet$ Conversely, when $\kappa$ is large, the qubit is able to leak information very fast to the outside world. However, this also makes it much less sensitive to the effect of being coupled, in addition to the outside world, to a HO with given photon number; in physicists' terms, the resonance peaks are broadened by the dissipation such that they cannot be well distinguished.
From a mathematical viewpoint, this regime allows to apply \emph{adiabatic elimination}, with approximation parameter $\chi/\kappa \ll 1$; instead of considering the system as essentially resonant on a particular coupling, this considers it as essentially dissipative, with the HO coupling treated as a perturbation (which precisely we want to detect!).
\newline Unsurprisingly, the most efficient regime to measure photon number is with $\chi/\kappa$ of order 1 (see $2\pi 4.9$ MHz vs. 20 MHz in the experiment). This cannot be faithfully covered by low-order expansions. In the present paper, by carrying out adiabatic elimination to arbitrary order, we essentially provide a solution that is valid for any value of $\chi/\kappa$ and thus provides correct (partial) information in all regimes.

Before proceeding, we make two straightforward and exact simplifications on the full system model. First, the \emph{expected measurement rate achieved by the setup} is bounded by the dissipation induced on the complementary variables by the term in $\mathcal{D}_L(\rho)$. We can therefore discard the stochastic terms and the output equation, to focus on the induced dissipation in a deterministic system. Second, because the only operator acting on the HO is the Hamiltonian $\mathbf{a}^\dagger \mathbf{a}$, the system naturally decouples: any part of the state spanned by a subset of eigenvectors of $\mathbf{a}^\dagger \mathbf{a}$ undergoes an autonomous evolution. The elementary building block is thus to take two eigenvectors, i.e.~focus on distinguishing between two photon numbers $n_1$ and $n_2$. The model describing this part of the dynamics, and studied in Section \ref{sec:result}, is:
\begin{eqnarray}\label{eq:model}
\tfrac{d}{dt}\rho & = &-i \left[\frac{\Omega}{2} \sigma_x + \frac{\tilde{\Delta}}{2}\sigma_z - \frac{\tilde{\chi}}{2}\tilde{\sigma}_z\otimes \sigma_z,\;\rho\right] + \kappa \mathcal{D}_{\sigma_-}(\rho) \; ,
\end{eqnarray}
where $\rho$ is the state on a Hilbert space $\mathbb{C}^2 \otimes\mathbb{C}^2$ equivalent to two qubits --- the one used for measurement, which we will call the \emph{measurement} qubit, and the effective one spanned by the two photon numbers to be distinguished, which we will call the \emph{target} qubit; $\tilde{\Delta} = \Delta - \frac{n_1+n_2}{2} \chi$ is the drive detuning with respect to the measurement qubit when the HO is centered in the middle between $n_1$ and $n_2$ photons; $\tilde{\chi} = \frac{(n_1-n_2)\chi}{2}$ expresses the effective coupling strength with the two photon numbers considered, using the tilde to distinguish the Pauli operator acting on the HO component. The goal is to eliminate the measurement qubit (all operators without tilde) and give an effective reduced model capturing the average dissipation induced by the leakage of information out of the HO component.


\subsection{The center manifold approach to adiabatic elimination}

Consider a general quantum system of the form
$$\tfrac{d}{dt} \rho = \mathcal{L}_0(\rho) + \epsilon \mathcal{L}_1(\rho) \; ,$$
where $\mathcal{L}_0$ and $\mathcal{L}_1$ are both superoperators of the type $-i[H,\rho] + \sum_k \mathcal{D}_{L_k}(\rho)$. We assume that the behavior of $\mathcal{L}_0$ is easy to analyze and makes $\rho$ converge towards a manifold of stable equilibria $\mathcal{S} = \{ \bar{\rho} \vert \mathcal{L}_0(\bar{\rho}) = 0 \}$. The goal is to express how this set gets perturbed by the presence of $\mathcal{L}_1$. Center manifold theory ensures that, for $\epsilon \ll 1$, there exists a manifold of same dimension as $\mathcal{S}$, $\epsilon$-close to $\mathcal{S}$, and on which the dynamics is $\epsilon$-slow (\cite{Fenichel79}). To compute both the manifold and the dynamics, a series expansion in $\epsilon$ can be used (\cite{carr-book}). For quantum systems, since $\mathcal{L}$ is linear in $\rho$, the manifolds boil down to subspaces and the variations due to $\mathcal{L}_1$, treated in bulk on the slow eigenspace which is separated from the fast converging one, should be analytic. The meaningful state space for quantum systems however is not linear, and this requires more care.

More precisely, we want to assign a quantum state $\rho_s$ to the reduced model, where for $\epsilon=0$, $\rho_s$ spans $\mathcal{S}$ and moves as $\tfrac{d}{dt}\rho_s = 0$. For $\epsilon \neq 0$, we search for:
\begin{itemize}
\item reduced dynamics of the form $\tfrac{d}{dt}\rho_s = \mathcal{L}_s(\rho_s) = -i[H_s,\rho] + \sum_k \mathcal{D}_{L_{s,k}}(\rho)$;
\item an embedding $\rho = \mathcal{K}(\rho_s) = \sum_k M_k \rho_s M_k^\dagger$ of $\rho_s$ into the full system, with $\sum_k M_k^\dagger M_k = I$; this form is a completely positive trace-preserving map, also called a Kraus map.
\end{itemize}
This allows us to analyze $\rho_s$ like a usual quantum system, while the associated $\rho$ remains physically meaningful.

To compute $\mathcal{L}_s$ and $\mathcal{K}$ we just impose invariance of the resulting subsystem under the actual dynamics:
\begin{equation}\label{eq:invariance}
\mathcal{K}(\mathcal{L}_s(\rho_s)) = (\mathcal{L}_0+\epsilon \mathcal{L}_1)(\mathcal{K}(\rho_s))\;\;\text{ for all } \rho_s \; .
\end{equation}
Since solving this exactly can be difficult, one can resort to a series expansion in $\epsilon$ and separately solve terms of different orders, increasing the power of $\epsilon$ to improve the accuracy of the approximation (\cite{carr-book}). Existing work has done this up to $\epsilon^2$ for general composite quantum systems (\cite{azouitQST2017,bouten-silberfarb-08,zanardi2016dissipative}), with $\mathcal{L}_0$  acting only on a fast subsystem and $\epsilon \mathcal{L}_1$ denoting its coupling to the subsystem essentially modeled by $\rho_s$. The result at this order has been explicitly put into quantum structure (\cite{azouitQST2017}), proving positivity preservation. For higher orders, formulas and proofs appear to get significantly more complex. Our goal is to carry both the formulas and the positivity proof to arbitrary order on the system \eqref{eq:model} and derive lessons from this.


\section{Adiabatic Elimination to arbitrary order}\label{sec:result}

In the system \eqref{eq:model}, we consider $\epsilon = -\tilde{\chi}/2$ and identify the corresponding form $\mathcal{L}_0 + \epsilon \mathcal{L}_1$ in \eqref{eq:invariance}. In particular, $\mathcal{L}_0 = \text{I} \otimes \bar{\mathcal{L}}$ only acts on the measurement qubit, with 
$\bar{\mathcal{L}}(\rho_m) := -i [\frac{\Omega}{2} \sigma_x + \frac{\tilde{\Delta}}{2}\sigma_z,\;\rho_m] + \kappa \mathcal{D}_{\sigma_-}(\rho_m)$, and $\mathcal{L}_1 = -i[\tilde{\sigma}_z\otimes \sigma_z,\;\rho]$ takes the form of an interaction Hamiltonian. 


\subsection{Formulas computing the reduced model}

We start by considering the series expansion of the adiabatic approximation. We denote by a superscript the contribution of a given order of approximation to $\mathcal{L}_s$ and $\mathcal{K}$, i.e.~$\mathcal{L}_s = \sum_{k=0}^{+\infty} \epsilon^k \mathcal{L}_s^{(k)}$ and $\mathcal{K} = \sum_{k=0}^{+\infty} \epsilon^k \mathcal{K}^{(k)}$. Note that to preserve the trace at all orders, we need $\text{tr}(\mathcal{K}^{(0)}(\rho_s)) = 1$ and $\text{tr}(\mathcal{K}^{(k)}(\rho_s)) = 0$ for all $k>0$.

To zero order ($\epsilon=0$), the target qubit undergoes no dynamics and the measurement qubit converges to the unique steady state $\bar{\rho}$ satisfying  $\bar{\mathcal{L}}(\bar{\rho}) = 0$. Thus to zero order, $\tfrac{d}{dt} \rho_s = \mathcal{L}_s^{(0)}(\rho_s)= 0$ and $\mathcal{K}^{(0)}(\rho_s) = \rho_s \otimes \bar{\rho}$.

The explicit solutions for the first two orders of approximation are corollaries of \cite{azouitQST2017}.
\vspace{2mm}

\noindent \textbf{Proposition 1, \cite{azouitQST2017}:} 
\emph{Solving the invariance equations at orders $\epsilon=\tfrac{-\tilde{\chi}}{2}$ and $\epsilon^2$ yields the reduced model:}
\begin{eqnarray*}
\tfrac{d}{dt}\rho_s &=& -i \epsilon c_1 [\tsz,\rho_s] + \epsilon^2 c_2 \mathcal{D}_{\tsz}(\rho_s) + O(\epsilon^3) \; ,\\
\mathcal{K}(\rho_s) &=& \rho_s \otimes \bar{\rho}\;\; - i\epsilon (\tsz \rho_s \otimes M_1 - \rho_s \tsz \otimes M_1^\dagger)\\
& & + \epsilon^2 (\tsz \rho_s \tsz \otimes M_2 - \rho_s \otimes M_0)  + O(\epsilon^3)\\
& & + (\epsilon \beta_1 + \epsilon^2 \beta_2) (\tsz \rho_s \tsz -\rho_s) \otimes \bar{\rho}
\end{eqnarray*}
\emph{where $c_1 = \text{tr}(\sigma_z \bar{\rho})$; $\; c_2= \text{tr}(\sigma_z (M_1+M_1^\dagger)) \geq 0$; $M_1$ is computed as the solution with $\text{tr}(M_1)=0$ of:}
$$-\bar{\mathcal{L}}(M_1) = \sigma_z \bar{\rho}-\text{tr}(\sigma_z \bar{\rho})\, \bar{\rho} \; ;$$
\emph{$M_0$ and $M_2$ are respectively computed as solutions of:}
\begin{eqnarray*}
\bar{\mathcal{L}}(M_2)&=&(M_1 \sigma_z + \sigma_z M_1^\dagger) + c_2 \bar{\rho} - c_1(M_1+M_1^\dagger) \\
\bar{\mathcal{L}}(M_0)&=&(\sigma_z M_1 + M_1^\dagger \sigma_z) + c_2 \bar{\rho} - c_1(M_1+M_1^\dagger)
\end{eqnarray*}
\emph{with $\text{tr}(M_0)=\text{tr}(M_2)=0$; and $\beta_1,\beta_2$ are free real parameters. Moreover, for every $\bar{\rho}$ full rank, i.e.~as soon as $\tilde{\Omega} \neq 0$, there exist $\beta_1,\beta_2,\epsilon$ making the approximate $\mathcal{K}$ completely positive.}
\vspace{2mm}

\noindent \emph{Proof:} The terms of first order in $\epsilon$ from \eqref{eq:invariance} give:
$$
\mathcal{L}_s^{(1)}(\rho_s) \otimes \bar{\rho} = \bar{\mathcal{L}}(\mathcal{K}^{(1)}(\rho_s)) - i[\tilde{\sigma}_z\otimes \sigma_z,\; \rho_s \otimes \bar{\rho}]\; .$$
Taking partial trace over the measurement qubit gives 
$\mathcal{L}_s^{(1)}(\rho_s) = -i\text{tr}(\sigma_z \bar{\rho})\; [\tilde{\sigma}_z,\, \rho_s]$. 
Plugging this back into the first-order condition, together with the proposed form of $\mathcal{K}^{(1)}$, yields the equation involving $M_1$. Its solution is discussed in \cite{azouitQST2017} and the fact that $\mathcal{L}(\bar{\rho})=0$ implies that the general solution can contain the term in $\beta_1$ (and others).
\newline At second order we repeat the procedure. Gathering terms of order $\epsilon^2$ from \eqref{eq:invariance}, we have
\begin{multline*}
\mathcal{L}_s^{(2)}(\rho_s) \otimes \bar{\rho} + \mathcal{K}^{(1)}(\mathcal{L}_s^{(1)}(\rho_s))\\ = \bar{\mathcal{L}}(\mathcal{K}^{(2)}(\rho_s)) - i[\tilde{\sigma}_z\otimes \sigma_z,\; \mathcal{K}^{(1)}(\rho_s)]\; .
\end{multline*}
Note that here the terms involving $\mathcal{K}^{(1)}$ contain, on the target qubit, operations of type $\tilde{\sigma_z} \rho_s \tilde{\sigma_z}$ or of type $(\tilde{\sigma_z})^2 \rho_s = \rho_s (\tilde{\sigma_z})^2 = \rho_s$. As before a partial trace allows to eliminate the term in $\mathcal{K}^{(2)}$ and write the explicit expression of $\mathcal{L}_s^{(2)}$; positivity of $c_2$ can be checked as in \cite{azouitQST2017}. The expressions for $\mathcal{K}^{(2)}$ are similarly obtained after plugging its form and the just computed $\mathcal{L}_s^{(2)}$ back into the invariance equation.\newline
The proposed $\mathcal{K}(\rho_s)$ with $\beta_1=0$ can be rewritten as e.g.:
\begin{eqnarray*}
\mathcal{K}(\rho_s) &=& \tfrac{1}{2}(I+i\epsilon \tsz \otimes 2 M_1) \rho_s (I-i\epsilon \tsz \otimes 2 M_1^\dagger) \\
&& + \epsilon^2 \tsz \rho_s \tsz \otimes (M_2 - 2 M_1 M_1^\dagger + \beta_2 \bar{\rho}) \\
&& + \rho_s \otimes (\tfrac{1-2\epsilon^2 \beta_2}{2} \bar{\rho} - \epsilon^2 M_0) \; .
\end{eqnarray*}
The first line is a positive expression; for given $\bar{\rho}$ of full rank, we can choose $\beta_2$ to make the second line positive, then $\epsilon$ to make the last line positive. We then obtain the explicit expression 
$\mathcal{K}(\rho_s) = \sum_j K_j \rho_s K_j^\dagger$ of a completely positive map.
\hfill $\square$\\

The higher order iterations turn out to follow a simple structure.
\vspace{2mm}

\noindent \textbf{Proposition 2:}\newline \emph{At any odd order $k$: $\mathcal{L}_s^{(k)}(\rho_s) = -i\,f_k [\tilde{\sigma}_z,\, \rho_s]$ for some real constant $f_k$ and $\mathcal{K}^{(k)}(\rho_s) = \tilde{\sigma}_z \rho_s \otimes M_{(k)} + \rho_s \tilde{\sigma}_z  \otimes M_{(k)}^\dagger$ for some operator $M_{(k)}$ on the measurement qubit.
\newline
At any even order $k$: $\mathcal{L}_s^{(k)}(\rho_s) = \frac{g_k}{2} (\tilde{\sigma}_z \rho_s \tilde{\sigma}_z - \rho_s)$ for some real constant $g_k$, and $\mathcal{K}^{(k)}(\rho_s) = \tilde{\sigma}_z \rho_s \tilde{\sigma}_z \otimes M^{(2)}_{(k)} - \rho_s \otimes M^{(0)}_{(k)}$ for some Hermitian operators $M^{(0)}_{(k)}, M^{(2)}_{(k)}$ on the measurement qubit.}
\vspace{2mm}

\noindent \emph{Proof:} We proceed by iteration. The property is true for $k=1,2$. The statement essentially holds because when plugging in all the knowledge from previous orders, the invariance condition for odd $k$ takes the same form as for $k=1$, while for even $k$ it takes the same form as for $k=2$. Indeed, for a general $k$, the invariance condition reads:
\begin{eqnarray*}
&&\mathcal{L}_s^{(k)}(\rho_s) \otimes \bar{\rho}
+{\textstyle \sum_{j=1}^{k-1}} \mathcal{K}^{(k-j)} (\mathcal{L}_s^{(j)}(\rho_s))\\
&&= \bar{\mathcal{L}}(\mathcal{K}^{(k)}(\rho_s)) - i[\tilde{\sigma}_z\otimes \sigma_z,\; \mathcal{K}^{(k-1)}(\rho_s)]\; .\end{eqnarray*}
The partial trace over measurement qubit gives an expression for $\mathcal{L}_s^{(k)}$. Assume that our form holds up to $k-1$.
\newline $\bullet$ For $k$ odd, thanks to $(\tsz)^2 \rho_s = \rho_s (\tsz)^2 = \rho_s$, each term on the left hand side contains a linear combination of $\tilde{\sigma}_z \rho_s$ and $\rho_s \tilde{\sigma}_z$ only. The same holds true for the remaining term $\text{tr}_{meas.qubit}(i[\tilde{\sigma}_z\otimes \sigma_z,\; \mathcal{K}^{(k-1)}(\rho_s)])$ on the right hand side. It is not hard to check that, provided our form holds true up to $k-1$, the terms in $\tilde{\sigma}_z \rho_s$ and in $\rho_s \tilde{\sigma}_z$ have opposite imaginary coefficient, confirming the form of $\mathcal{L}_s^{(k)}$. For instance, for $j=2$ on the left, we have
\begin{eqnarray*}
&& \tilde{\sigma}_z \rho_s \otimes M_{k-2} + \rho_s \tilde{\sigma}_z \otimes M_{k-2}^\dagger \\
&& - \tilde{\sigma}_z(\,\tilde{\sigma}_z \rho_s \otimes M_{k-2} + \rho_s \tilde{\sigma}_z \otimes M_{k-2}^\dagger  \,) \tilde{\sigma}_z \\
&& = [\sigma_z,\rho_s] \otimes (M_{k-2} - M_{k-2}^\dagger)
\end{eqnarray*}
and $\text{tr}(M_{k-2} - M_{k-2}^\dagger)$ is imaginary.
\newline $\bullet$ Similarly, for $k$ even, each term on the left hand side contains a linear combination of $\tilde{\sigma}_z \rho_s \tilde{\sigma}_z$ and $\rho_s$ only, as does the remaining term on the right hand side; and when actually checking a term it is obvious that the coefficients of $\tilde{\sigma}_z \rho_s \tilde{\sigma}_z$ and of $\rho_s$ are real and opposite.
\newline $\bullet$ This form of the equations also implies the same type of solution for $\mathcal{K}^{(k)}$, i.e.~$k$ even is like $k=2$ and $k$ odd is like $k=1$. Note that we do not claim here (yet) to ensure $g_k$ positive or $\mathcal{K}$ completely positive, so we have nothing more to prove.\hfill $\square$\\

The series expansion from Proposition 2 can also be summarized in the following form:
\begin{eqnarray}\label{eq:fullform}
\mathcal{L}_s(\rho_s) &=& -i f [\tsz,\,\rho_s] + \frac{g}{2} \mathcal{D}_{\tsz}(\rho_s) \\
\nonumber \mathcal{K}(\rho_s) &=& \tfrac{(I\text{+}\tsz)}{2} \rho_s \tfrac{(I\text{+}\tsz)}{2} \otimes Q_0 + \tfrac{(I\text{-}\tsz)}{2} \rho_s \tfrac{(I\text{-}\tsz)}{2} \otimes Q_1\\
\nonumber && + \tfrac{\tsz \rho_s \tsz-\rho_s}{2} \otimes Q_2 + \tfrac{i(\tsz \rho_s - \rho_s \tsz)}{2} \otimes Q_3
\end{eqnarray}
where $f$ is odd in $\epsilon$; $g$ is even in $\epsilon$ and still must be proven positive; $Q_k$ for $k=0,1,2,3$ are Hermitian operators. To satisfy trace preservation i.e.~$\text{tr}(\mathcal{K}(\rho_s))=1$ for all $\rho_s$, we need $\text{tr}(Q_0)=\text{tr}(Q_1)=1$. We further must prove complete positivity of this form, which we will do in the next subsection. We first set out to formally solve the system as a whole.

Note that the first line of $\mathcal{K}$ in \eqref{eq:fullform} extracts only the diagonal components of $\rho_s$ in the $\tsz$ basis, while the second line extracts only the off-diagonal components.
\vspace{2mm}

\noindent \textbf{Theorem 3:} \emph{The system \eqref{eq:model} admits an invariant subsystem of the form \eqref{eq:fullform}, with:}
\begin{itemize}
\item $Q_0 = \bar{\rho}_0$ \emph{the steady state of the measurement qubit if the target qubit was in the ground state $\rho_s = \frac{(I+\tsz)}{2}$; this state satisfies $\bar{\mathcal{L}}(\bar{\rho}_0) +i\frac{\tilde{\chi}}{2}[\sigma_z, \bar{\rho}_0]=0$.}
\item $Q_1 = \bar{\rho}_1$ \emph{the steady state of the measurement qubit if the target qubit was in the excited state $\rho_s = \frac{(I-\tsz)}{2}$; this state satisfies $\bar{\mathcal{L}}(\bar{\rho}_1) -i\frac{\tilde{\chi}}{2}[\sigma_z, \bar{\rho}_1]=0$.}
\item \emph{$g = \frac{-(\lambda + \lambda^*)}{2} \geq 0$ and $f = \frac{\lambda - \lambda^*}{4i}$, where $\lambda$ is an eigenvalue of the matrix}
$$A = \left(\begin{array}{cccc}
-\kappa/2 & -\tilde{\Delta} & 0 & 0 \\
\tilde{\Delta} & -\kappa/2 & -\Omega & 0 \\
0 & \Omega & -\kappa & -\kappa - i \tilde{\chi} \\
0 & 0 & -i \tilde{\chi} & 0 
\end{array}\right) \; .$$ 
\item \emph{$Q_2= c \tilde{Q}_2$ and $Q_3 = c \tilde{Q}_3$ where the vectorized versions of $\tilde{Q}_2$ and $\tilde{Q}_3$ correspond respectively to the real part and imaginary part of the eigenvector of $A$ associated to $\lambda$, and $c$ is an arbitrary complex number.}
\end{itemize}
\vspace{2mm}

\noindent \emph{Proof:} Plugging the form \eqref{eq:fullform} into the invariance equation, we get expressions on the measurement qubit, multiplying factors of the form $\tsz \rho_s$, $\rho_s \tsz$, $\rho_s$ and $\tsz \rho_s \tsz$. Rearranging and separating these four terms yields:
\begin{eqnarray*}
&&\bar{\mathcal{L}}(Q_0) +i\tfrac{\tilde{\chi}}{2}[\sigma_z, Q_0] = 0\\
&&\bar{\mathcal{L}}(Q_1) -i\tfrac{\tilde{\chi}}{2}[\sigma_z, Q_1] = 0\\
&&\bar{\mathcal{L}}(Q_2) = - g Q_2 + 2i f (iQ_3) + i\tfrac{\tilde{\chi}}{2}(\,\sz (iQ_3)+(iQ_3)\sz\,)\\
&&\bar{\mathcal{L}}(iQ_3) = - g (iQ_3) + 2i f Q_2 + i\tfrac{\tilde{\chi}}{2}(\,\sz Q_2+Q_2\sz\,) \; .
\end{eqnarray*}
The first two equations characterize $Q_0$ and $Q_1$ up to a scalar factor; the conditions $\text{tr}(Q_0)=\text{tr}(Q_1)=1$ fix this scalar to yield the first part of the solution.
\newline Defining $S_+ = Q_2 + iQ_3$ and $S_- = Q_2-iQ_3 = (S_+)^\dagger$, the remaining two equations decouple into
$$ \bar{\mathcal{L}}(S_+) -i \tfrac{\tilde{\chi}}{2} (\sz S_+ + S_+ \sz) = - (g-2if) S_+ $$
and its hermitian conjugate. Since $g$ and $f$ are part of the unknowns, we here have an eigenvalue equation on the Hilbert space of the measurement qubit. The eigenvalues give the reduced dynamics -- real part for $g$, imaginary part for $f$ -- and the eigenvectors give $Q_2$, $Q_3$. The matrix $A$ in the statement corresponds to parameterizing $S_+ = \alpha_1 \sigma_x + \alpha_2 \sigma_y + \alpha_3 \sz + \alpha_4 I$, with complex coefficients $\alpha_1$ to $\alpha_4$ stacked in this order into a column vector. Since we have no further equality conditions, the eigenvector is defined up to a scalar factor, $c$ in the statement. Matrix $A$ is rigorously the state matrix governing the evolution of a component $\q{n_2}\qd{n_1} \otimes Q(t)$ of the full quantum system; the latter cannot be unstable so $A$ must bes stable, ensuring that $g$ is positive. This can also be checked, although a bit tediously, with the generalized Routh-Hurwitz criterion. \hfill $\square$ \\

\emph{Computational efficiency of model reduction:} Compared to studying the full system, we now have an eigenvalue equation on the Hilbert space of the measurement qubit only. Compared to a finite order expansion, we now solve an eigenvalue equation for the measurement qubit dynamics, instead of computing the inverse of this dynamics.

\emph{Regarding eigenvalues:} Matrix $A$ in Thm.3 has 4 eigenvalues, yielding 4 possible reduced dynamics. For $\tilde{\chi}$ small, a single eigenvalue is close to zero, giving the slow dynamics that we search to characterize. For larger $\tilde{\chi}$, the 4 eigenvalues become of similar order. One may want to select the eigenspace which follows analytically from $\tilde{\chi}=0$, with the idea of recovering Prop.2. Alternatively, one may acknowledge that we are really interested in the evolution of $\text{tr}_{meas.qubit}(\rho)$, corresponding essentially to $\alpha_4$ in the proof of Thm.3. Generically, $\alpha_4$ does not follow autonomous dynamics, it is governed by all 4 eigenvalues of $A$, and on the long run the slowest one will dominate. From this viewpoint, the effective measurement rate is obtained by taking $g$ corresponding to the slowest eigenvalue of $A$.

\emph{Regarding eigenvectors:} The free scalar factor $c$ in the statement indicates a gauge degree of freedom on how $\rho_s$ is mapped into the full space; e.g.~we could a priori use $\rho'_s = U \rho_s U^\dagger$ instead of $\rho_s$, with some arbitrary unitary $U$, write the dynamics $\mathcal{L}'_s$ on $\rho'_s$, and map this to the full space as $\mathcal{K}'(\rho'_s) = \mathcal{K}(U^\dagger \rho'_s U)$. By imposing the form with $\tsz$, we are taking away some gauge freedom, but not all. In particular, if $\mathcal{K}$ is the map from Thm.3 associated to a scalar $c$ and $\mathcal{K}'$ is the map associated to $c' = e^{i\theta}\, c$, one checks that this corresponds to $\rho'_s = e^{i\theta\sigma_z/2} \rho_s e^{-i\theta\tsz/2}$. This makes physical sense as the dynamics, involving $\tsz$ only, is indeed invariant under this unitary basis transformation. It is most natural to assume $\mathcal{K}(\rho_s) = \rho_s \otimes \bar{\rho}$ for $\epsilon=0$, but this still leaves the choice to define $\rho'_s$ with an arbitrary function $\theta(\epsilon)$ satisfying $\theta(0)=0$.

Changing $|c|$ also has a clear effect, in relation with ensuring complete positivity of $\mathcal{K}$, as we discuss next.


\subsection{Ensuring complete positivity}\label{ssec:positivity}

The norm of $c$ defining $Q_2$ and $Q_3$ in Thm.3 determines whether an off-diagonal element of $\rho_s$ will be mapped to a small or large contribution in the actual $\rho$. For the extreme case $c=Q_2=Q_3=0$, the off-diagonals of $\rho_s$ would be in the kernel of $\mathcal{K}$ 
 and the reduced model would just describe that a state of the form $p\, \q{n_1}\qd{n_1} \otimes \bar{\rho}_0 + (1-p) \q{n_2}\qd{n_2} \otimes \bar{\rho}_1$ with $p \in [0,1]$ is a steady state of the full dynamics. Taking $c\neq 0$ allows to model the target qubit coherences, i.e.~terms involving $\q{n_1}\qd{n_2}$. On the other hand, taking $c$ very large, the result of $\mathcal{K}(\rho_s)$ dominated by such off-diagonal terms would not be positive i.e.~not a proper quantum state. We recall that $\rho$ must be a positive-semidefinite Hermitian matrix (we just say ``positive'') of trace $1$ in order to represent a quantum state. We have the following remarkable result.
\vspace{1mm}

\noindent \textbf{Theorem 4:} \emph{There exists a value for the scalar factor $c$ in Theorem 3 such that:}
\begin{itemize}
\item \emph{the resulting map $\mathcal{K}$ is positive on the qubit state space, i.e.~it maps every positive $\rho_s$ to positive $\rho$, and for any $c' > c$ this would not be true anymore;}
\item \emph{the resulting map $\mathcal{K}$ is completely positive, i.e.~$\mathcal{K} \otimes I$ is positive on the state space of two qubits, and for any $c' > c$ this would not be true anymore;}
\item \emph{the resulting map $\mathcal{K}$ covers the full invariant manifold, i.e.~every positive $\rho$ on the invariant subspace takes the form $\mathcal{K}(\rho_s)$ for some positive $\rho_s$.}
\end{itemize}

\noindent \emph{Proof:} The key ingredient of our proof is the Schur complement argument: block matrix $[A,\; B\; ; \; B^\dagger,\; C]$ is positive if and only if $C$ is positive and $A-BC^{-1}B^\dagger$ is positive.

Writing $\rho_{n,m} = \qd{n} \rho_s \q{m}$, we can rewrite \eqref{eq:fullform} in the form:
\begin{eqnarray*}
\mathcal{K}(\rho_s) &=& \rho_{n_1,n_1}\, \q{n_1}\qd{n_1} \otimes \bar{\rho}_0 + \rho_{n_2,n_2}\, \q{n_2}\qd{n_2} \otimes \bar{\rho}_1 \\
&& + \rho_{n_1,n_2}\, \q{n_1}\qd{n_2} \otimes M + \rho_{n_2,n_1}\, \q{n_2}\qd{n_1} \otimes M^\dagger \; .
\end{eqnarray*}
\emph{Positivity:} Applying the Schur argument to the blocks distinguished by $n_1,n_2$, map $\mathcal{K}$ is positive if and only if
\newline 1. $\rho_{n_2,n_2}\bar{\rho}_1$ is positive (OK whenever $\rho_s$ is positive); and
\newline 2. $\rho_{n_1,n_1} \bar{\rho}_0 - \frac{\rho_{n_1,n_2} \rho_{n_2,n_1}}{\rho_{n_2,n_2}} M \bar{\rho}_1^{-1} M^\dagger$ is positive. This depends on $M$. The positive $\rho_s$ can span all cases with $\rho_{n_1,n_1} \geq \rho_{n_1,n_2} \rho_{n_2,n_1} / \rho_{n_2,n_2}$, in particular with equality. Thus, considering the worst case, we must have
\begin{equation}\label{eq:poscond}
\bar{\rho}_0 \geq M \bar{\rho}_1^{-1} M^\dagger \; ,
\end{equation}
in other words matrix $[\bar{\rho}_0,\; M\; ; \; M^\dagger,\; \bar{\rho}_1]$ positive. The first part of the statement just amounts to selecting the value of $c$ saturating this condition.

\noindent \emph{Complete positivity:} A positive map acting on a qubit, is by definition completely positive if $\mathcal{K} \otimes I$ is positive on the state space of two qubits. In matrix form, we need
\begin{equation}\label{eq:ForSchur}
[\mathcal{K}(A),\; \mathcal{K}(B)\; ; \; \mathcal{K}(B^\dagger),\; \mathcal{K}(C)]
\end{equation}
positive for any positive state $\hat{\rho} = [A,\; B\; ; \; B^\dagger,\; C]$ of two qubits, meaning with $A,B,C$ being operators on $\mathbb{C}^2$ and $\hat{\rho}$ satisfying the Schur conditions. Applying the Schur argument to \eqref{eq:ForSchur} with our expression of $\mathcal{K}$ and using tensor product properties, we get the conditions:
\newline 1. $C \otimes \bar{\rho}_1$ must be positive. Since $C$ must be positive for $\hat{\rho}$ to be positive, this is always satisfied as $\bar{\rho}_1$ is positive.
\newline 2. $A \otimes \bar{\rho}_0 - (B^\dagger\, C^{-1}\, B) \otimes (M \bar{\rho}_1^{-1} M^\dagger)$ must be positive. Since $\{ X>\tilde{X}$ and $Y>\tilde{Y} \}$ is sufficient to imply $X \otimes Y > \tilde{X} \otimes \tilde{Y}$, the condition is satisfied when $\hat{\rho}$ is positive (whose Schur argument yields the $X>\tilde{X}$ property) and $\bar{\rho}_0 > M \bar{\rho}_1^{-1} M^\dagger$ (the $Y>\tilde{Y}$ property). The latter is the condition \eqref{eq:poscond} for positivity of $\mathcal{K}$, proving the second point.

\noindent \emph{Surjectivity:} We must show that if $\mathcal{K}(q)$ is positive (of trace one), then $q$ is positive (of trace one), for the Kraus map saturating the condition \eqref{eq:poscond}. By the Schur argument, having $\mathcal{K}(q)$ positive first requires $q_{n_2,n_2} \bar{\rho}_1$ positive and thus $q_{n_2,n_2}\geq 0$; reversing the roles of $n_1$ and $n_2$ in this argument, we get the condition $q_{n_1,n_1} \geq 0$. The remaining Schur condition for positivity of $\mathcal{K}(q)$ is
$q_{n_1,n_1}\, \bar{\rho}_0 - \frac{q_{n_1,n_2} q_{n_2,n_1}}{q_{n_2,n_2}}\, M \bar{\rho}_1^{-1} M^\dagger$ positive, or equivalently 
\begin{equation}\label{eq:proofl}
\bar{\rho}_0 - r\, M \bar{\rho}_1^{-1} M^\dagger
\end{equation}
positive where $r=\frac{q_{n_1,n_2} q_{n_2,n_1}}{q_{n_2,n_2} q_{n_1,n_1}}$. Having $c$ saturate the condition \eqref{eq:poscond} means that $\bar{\rho}_0 - \frac{|\tilde{c}|^2}{|c|^2}\, M \bar{\rho}_1^{-1} M^\dagger$ cannot be positive for $\frac{|\tilde{c}|}{|c|}>1$. Thus in \eqref{eq:proofl} we must take $r\leq 1$, which implies that $q$ is positive. \hfill $\square$\\


\section{Analyzing the reduced system}\label{sec:conseq}

From this model reduction, we can deduce fundamental properties about the measurement setup performance.

\noindent \emph{Information-theoretic bound:} It should be impossible to acquire information about the photon number in the target system, faster than information is leaking out of the whole setup through the measurement qubit. The output signal leaking out of the latter is proportional to $\kappa\, \text{tr}(\rho \sigma_x)$, while the (indirect) output signal corresponding to \eqref{eq:fullform} would be $g\, \text{tr}(\rho \sigma_z)$. (This is why we have used $g/2$ instead of $g$ in \eqref{eq:fullform}.) Matrix $A$ in Thm.3 has only stable eigenvalues, with their sum $\text{tr}(A)=-2 \kappa$, so we readily see that the dissipation rate in the reduced model satisfies $g/2 = -(\lambda + \lambda^*)/4 \leq \kappa$ even when taking the fastest eigenvalue. Applying the generalized Routh-Hurwitz criterion allows to tighten this, proving $g/2 \leq \kappa/2$. Asymptotically, we would rather expect to be bounded by the slowest eigenvalue, which is readily bounded by $\text{tr}(A)/4$ and thus $g/2 \leq \kappa/4$.

\noindent \emph{Optimal operation point:} To optimize induced measurement rate, i.e.~$g$ in Thm.3, we can just study the slowest eigenvalue of $A$ as function of the parameters $\Omega,\tilde{\Delta},\kappa$ while the coupling $\tilde{\chi}$ between target system and measurement device is just fixed to its highest possible value. From the intuition discussed in the introduction, we expect that $\kappa$ should take an intermediate value, $\Omega$ high enough to get $\bar{\rho}$ significantly different from ground state at this $\kappa$; regarding $\tilde{\Delta}$, it was not a priori clear what is best, as the resonance intuition ($\tilde{\Delta} = \pm\tilde{\chi}$) only holds for $\tilde{\Omega},\kappa$ small. The eigenvalues of $A$ can be investigated by root locus analysis as function of $\tilde{\Omega}^2$ or $\tilde{\Delta}^2$, with other parameters fixed. Exploring this root locus for various $\kappa$, the largest $g$ appears when $A$ has two equal eigenvalues, at $\tilde{\Delta}=0$. Plugging this condition into the root polynomial, we obtain $\frac{(\tilde{\Omega}^2 - \kappa^2/4 + \tilde{\chi}^2 + i \kappa \tilde{\chi})^3}{\tilde{\Omega}^4\kappa^2}= \frac{27}{16}$. This together fixes the operating point $\tilde{\Omega}^2 \simeq 5.6058... (\frac{\tilde{\chi}}{2})^2$, $\kappa \simeq 4.3055...\frac{\tilde{\chi}}{2}$ and enables a measurement rate $g\simeq 1.2424... \frac{\tilde{\chi}}{2}$. Local optimality of the degenerate situation can be understood as follows. When a situation with two equal eigenvalues is perturbed by a complex parameter $\delta$, generically the eigenvalues split as $\sqrt{\delta}$ and this is the dominating effect; except when the eigenvalues split along a purely imaginary direction, which for a generic \emph{complex} matrices is not typical, one of the two eigenvalues gets closer to $0$. Thus, moving away from the degenerate situation is less optimal on the worst eigenvalue, at least locally.

\subsection{Comparison to experimental results}

A more accurate model of the experimental setup includes the finite damping time of the harmonic oscillator:
\begin{eqnarray}\label{eq:expnow}
\tfrac{d}{dt}\rho &=& -i [\dfrac{-\chi}{2}a^\dag a \otimes \sigma_\mathrm{z}+\dfrac{\Delta}{2}\sigma_\mathrm{z}+\dfrac{\Omega}{2}\sigma_\mathrm{x},\rho] + \kappa \mathcal{D}_{\sigma_-}(\rho) \\
\nonumber  &&  + \Gamma_\mathrm{1} \mathcal{D}_{a}(\rho)+ 2\Gamma_\mathrm{\phi} \mathcal{D}_{a^\dag a}(\rho),
\end{eqnarray}
with small damping rates $\Gamma_1$,$\Gamma_\phi$. The ``dephasing'' term in $\Gamma_\phi$ perturbs the phase of the mode (conjugate variable to the photon number $\ba^\dagger \ba$), like the measurement, and its rate just adds up to the effect of the measurement computed by adiabatic elimination. The ``relaxation'' term in $\Gamma_1$ describes energy loss; it implies that photon number is not exactly conserved and the decoupling into two-dimensional subspaces, leading to \eqref{eq:model}, is not exact anymore.

In the experiments, the harmonic oscillator is initialized in a coherent state, which is a superposition of different photon numbers. Then the qubit is driven and continuously measured during a time $t$. To compare our predictions with the actual system state, at time $t$ a standard direct Wigner tomography measurement of the harmonic oscillator is performed with an auxiliary device. By repeating the experiment many times for the same parameter values, and for different end times $t$, this gives access to the density matrix of the harmonic oscillator $\rho_\mathrm{HO}(t) = tr_{meas.qubit}(\rho(t))$ corresponding to the evolution \eqref{eq:expnow}.

\begin{figure*}
    \centering
    \includegraphics[width=14cm]{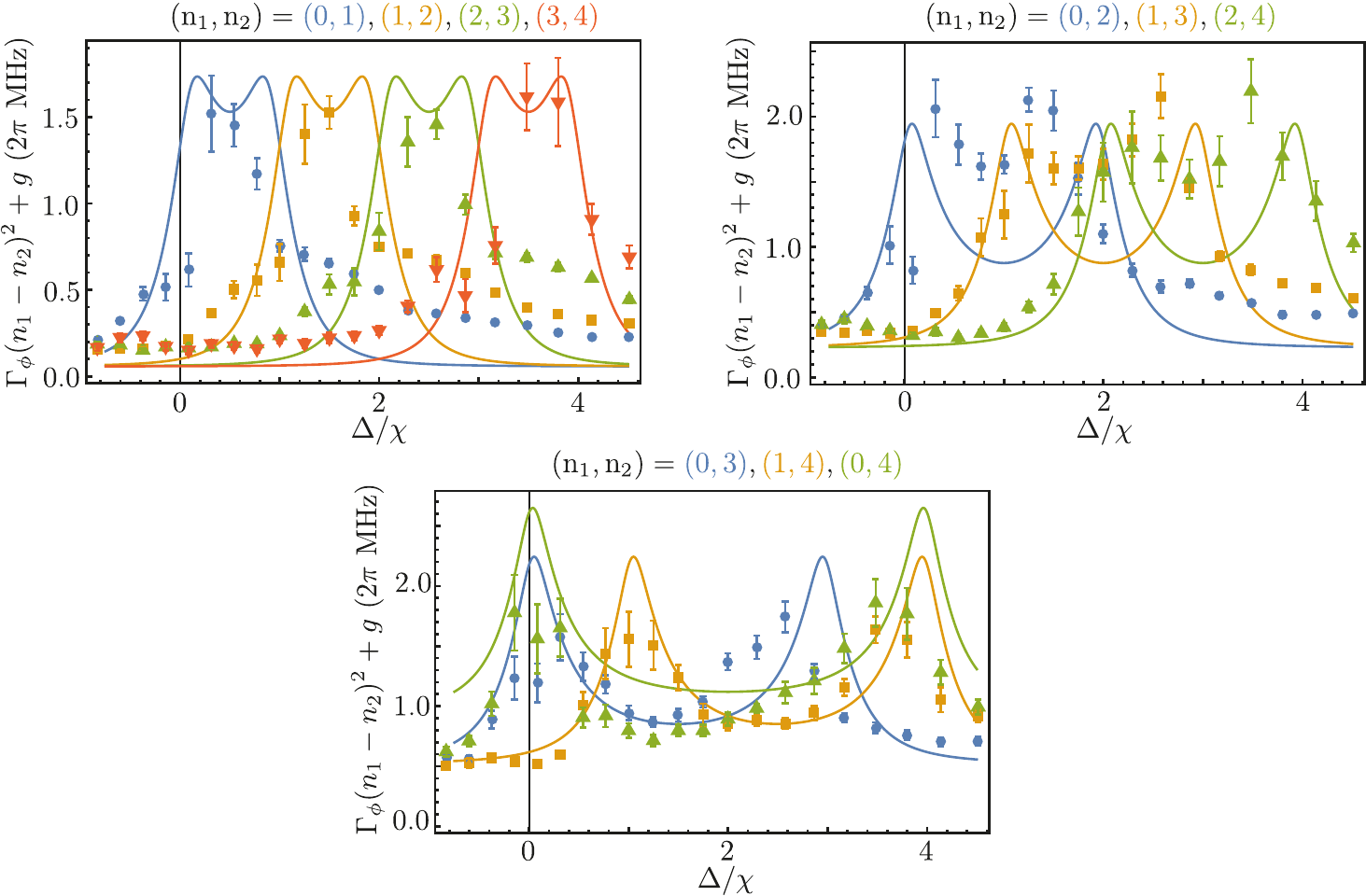}
    \caption{Measured decay rate of $ | \rho_{\mathrm{HO},n_1n_2} | / \sqrt{ \rho_{\mathrm{HO},n_1n_1} \rho_{\mathrm{HO},n_2n_2} } $ (dots) and adiabatic elimination theory (line) as a function of detuning $\Delta$ in units of $\chi$.}
    \label{fig:experience}
\end{figure*}

The impact of the continuous-time photon number measurement should be most characteristic on variables of the form 
$|\rho_{\mathrm{HO},n_1n_2}|/\sqrt{\rho_{\mathrm{HO},n_1n_1}\rho_{\mathrm{HO},n_2n_2}}$, where the indices $n_1,n_2$ denote components in photon number basis. Indeed, assuming $\Gamma_1=0$ and using the adiabatic elimination results, the dynamics for $\rho_\mathrm{s}$ in the subspace spanned by the states with $n_1$ or $n_2$ photons reads
\begin{equation}
    \dot{\rho_\mathrm{s}} = -if[\Tilde{\sigma}_\mathrm{z},\rho_\mathrm{s}]+\dfrac{g}{2}\mathcal{D}_{\Tilde{\sigma}_\mathrm{z}}(\rho_\mathrm{s})+ \dfrac{\Gamma_\mathrm{\phi}}{2}(n_1-n_2)^2 \mathcal{D}_{\Tilde{\sigma}_\mathrm{z}}(\rho_\mathrm{s}).
\end{equation}
This equation predicts an exponential decrease:
$$\dfrac{ |\rho_{\mathrm{s},n_1n_2}|}{\sqrt{\rho_{\mathrm{s},n_1n_1}\rho_{\mathrm{s},n_2n_2}}}(t) = e^{-(\Gamma_\phi (n_1-n_2)^2+g)t} \; .$$
On the experimental results, we do observe an exponential decrease of $|\rho_{\mathrm{HO},n_1n_2}|/\sqrt{\rho_{\mathrm{HO},n_1n_1}\rho_{\mathrm{HO},n_2n_2}}(t)$. We can extract the corresponding decay rate, to compare it with the prediction $\Gamma_\phi (n_1-n_2)^2+g$ of adiabatic elimination with $\Gamma_1=0$ and computing $g$ using the matrix $A$, for various values of $n_1,n_2$ and of the detuning $\Delta$. As shown on Figure \ref{fig:experience}, theory and experiment are in good agreement, without having to adjust any model parameters. The differences between the two are most probably due to the approximation $\Gamma_1=0$ in the theoretical model, while in reality $1/\Gamma_1 = 3.8~\mu \mathrm{s}$ for an experiment duration of about 5 $\mu$s.

%


\section{Conclusion}

We have been able to relate an adiabatic elimination series expansion at arbitrary order, to an exact model reduction in quantum form, for all parameter values on an existing experimental setup. The fact that the target system is a qubit subject to a single $\sigma_z$ coupling is the key to obtaining ideal results. However, we believe that Proposition 2 together with Theorems 3 and 4 indicate how high-order adiabatic elimination should behave more generally, at least for dispersive-type coupling (single term in the coupling Hamiltonian). Namely, there should exist simple gauge conditions ensuring a completely positive reduced model, and if not exactly surjective as in Thm.4, the model should at least be close to it. The resulting reduced model could miss states $\rho$ which would be small extrapolations of the image of $\mathcal{K}(\rho_s)$.

\bibliographystyle{plain}
\bibliography{BiblioIFAC}

\end{document}